# Multi-Period Do-Not-Exceed Limit for Variable Renewable Generation Dispatch Considering Discrete Recourse Controls

Zhigang Li, *Member, IEEE*, Feng Qiu, *Member, IEEE*, and Jianhui Wang, *Senior Member, IEEE*

*Abstract*—The do-not-exceed (DNE) limit method was proposed to accommodate more variable renewable generation (VRG) securely. However, the lack of involving discrete recourse control precludes this method from gaining more flexibility for better VRG integration. This letter formulates a multi-period DNE limit model considering continuous and discrete recourse controls. This model belongs to two-stage robust optimization with mixed integer recourse. A nested column-and-constraint generation approach is employed to solve this model. Case studies show the effectiveness of the proposed method.

*Index Terms*—do-not-exceed limit, flexibility, renewable energy, robust optimization.

## I. Introduction

Predetermined dispatch ranges are being adopted to achieve a full absorption of the renewable energy while maintaining the system security [1]-[2]. As one type of dispatchable ranges, do-not-exceed (DNE) limit is introduced to define the maximum output ranges of VRG that can be accommodated by the power grids reliably given the available regulation resources [1]. The DNE limit method proposed in [1] computes the single-period DNE limits of VRG considering continuous recourse controls, i.e., redispatch of unit generation output. The power dispatch framework based on the DNE limit is reported to outperform the conventional one in terms of VRG utilization and security guarantee.

However, in the current DNE approaches, only a single period is considered when determining the dispatchable ranges, which could result in sub-optimal or even infeasible dispatch ranges over a long run. Furthermore, discretely controllable devices, which can provide significant flexibility for corrective control in response to VRG uncertainty, are left out in DNE calculation due to the computational complexity caused by the discrete control variables.

To further explore the system flexibility, this letter investigates the multi-period DNE limit for VRG dispatch considering both continuous and discrete recourse controls (DRCs). The decision model for DNE limits is formulated as a two-stage robust optimization (RO) with mixed integer recourse. The introduction of DRCs makes this model more complicated to solve than conventional two-stage RO. The nested column-and-constraint generation (NC&CG) technique [4] is employed to solve the proposed model. The proposed approach broadens the maximum allowable ranges of VRG output by additionally exploiting available discrete regulation resources.

Z. Li is with the School of Electric Power Engineering, South China University of Technology, Guangzhou 510641, China (email: lizg16@scut.edu.cn). F. Qiu and J. Wang are with the Energy Systems Division, Argonne National Laboratory, Argonne, IL 60439 USA.

## II. Multi-Period DNE Limit with DRCs

The DNE limit approach seeks the maximum output ranges of VRG units, within which any VRG output can be handled by corrective controls to satisfy operation and security constraints. As a typical discretely controllable device, QSU is considered in this letter to provide DRC. Other DRCs such as responsive loads can also be included in the proposed model in a similar way. The proposed multi-period DNE limit problem is formulated in an abstract form as the following:

$$\max_{\{l_t, u_t, x_t(\cdot) \in \mathbb{R}^C, z_t(\cdot) \in \mathbb{N}^D, \forall t \in \mathcal{T}\}} \sum_{t \in \mathcal{T}} \sigma_t^T (u_t - l_t) \quad (1)$$

$$s.t. \quad w_t^{\min} \leq l_t \leq w_t^* \leq u_t \leq w_t^{\max}, \forall t \in \mathcal{T}, \quad (2)$$

$$\forall \tilde{w}_t \in [l_t, u_t], \exists x_t(\tilde{w}), z_t(\tilde{w}), t \in \mathcal{T}, s.t.:,$$

$$A_t x_t(\tilde{w}) + B_t z_t(\tilde{w}) + C_t \tilde{w}_t \leq d_t, \forall t \in \mathcal{T}, \quad (3)$$

$$\sum_{t \in \mathcal{T}} E_t x_t(\tilde{w}) + \sum_{t \in \mathcal{T}} F_t z_t(\tilde{w}) \leq g \quad (4)$$

where $l_t, u_t \in \mathbb{R}^w$ denote the lower and upper DNE limits of VRG, respectively. $x_t$ and $z_t$ represent the continuous (e.g., generation output of conventional units) and discrete (e.g., commitments of QSUs) resource actions, respectively.

The objective function in (1) is to maximize the weighted total VRG output ranges, where the coefficients $\sigma_t$ can be set using the strategy in [1]. $\mathcal{T}$ is the index set of periods. Equation (2) denotes the limit constraints for $l_t$ and $u_t$, where $w_t^{\min}$ and $w_t^{\max}$ denote the minimum and maximum capacities of VRG units. $w_t^*$ denotes the predicted output of VRG units. Equations (3)-(4) include the corrective dispatch constraints given any realization of VRG output $\tilde{w}_t$ within the DNE limits $[l_t, u_t]$. Equation (3) denotes the temporally decoupled constraints, such as the power balance, transmission capacity, and output limit constraints [1]. Equations (4) denote the temporally coupled constraints, such as ramping rates, commitment logics, and minimum up/down time constraints [3].

The proposed model is general enough to calculate the DNE limits with both continuous and discrete recourse controls. Particularly, when the index set $\mathcal{T}$ is a singleton and the discrete recourse variables $z_t(\cdot)$ are absent, the proposed model reduces to the original DNE limit model described in [1].

## III. Solution Method

The proposed model is described as a two-stage RO with mixed integer recourse as follows:

$$\max_{l, u, x(\cdot) \in \mathbb{R}, z(\cdot) \in \mathbb{N}} \sigma^T (u - l) \quad (5)$$

$$s.t. \quad w^{\min} \leq l \leq w^* \leq u \leq w^{\max}, \quad (6)$$

$$Hx(\tilde{w}) + Jz(\tilde{w}) \leq h - K\tilde{w}, \forall \tilde{w} \in [l, u], \quad (7)$$

where (7) denote the constraints in (3)-(4). The first stage determines the DNE limits of VRG output. The second stage checks the feasibility of corrective dispatch. The uncertain VRG output $\tilde{w}$ is parameterized using a normalized auxiliary variable $\tilde{v}$, i.e., $\tilde{w} = l + (u-l) \circ \tilde{v}$, where $\circ$ represents the component-wise product. Then (7) is rewritten as follows:

$$Hx(\tilde{v}) + Jz(\tilde{v}) \leq h - K[l + (u-l) \circ \tilde{v}], \forall \tilde{v} \in [0,1]^n \quad (8)$$

The proposed model can be solved accurately via the iterative NC&CG method [4]. The solution procedure is briefly described as follows (see [4] for details):

**Step 0**: Initialize. Set $k = 0$ and $K = \varnothing$.

**Step 1**: Solve the master problem. Solve the **MP** and obtain an optimal solution $l^{(k)}, u^{(k)}$.

$$\textbf{(MP)} \quad \max_{l,u,\{x^{(k)} \in \mathbb{R}, z^{(k)} \in \mathbb{N}: \forall k \in K\}} \sigma^T(u-l) \quad (9)$$

$$s.t. \text{ Equation (6),}$$
$$Hx^{(k)} + Jz^{(k)} + K[l + (u-l) \circ \tilde{v}^{(k)}] \leq h, \forall k \in K \quad (10)$$

**Step 2**: Solve the subproblem. Solve the **SP** and obtain an optimal solution $\tilde{v}^{(k+1)}$. If $Q(l^{(k)}, u^{(k)}) > 0$, set $K := K \cup \{k+1\}$ and $k := k+1$, and go to Step 1; Otherwise, stop the procedure.

$$\textbf{(SP)} \quad Q(l,u) = \max_{\tilde{v} \in [0,1]^n} \min_{s \geq 0, x \in \mathbb{R}, z \in \mathbb{N}} \mathbf{1}^T s = \max_{\tilde{v} \in [0,1]^n} \min_{z \in \mathbb{N}} \min_{s \geq 0, x \in \mathbb{R}} \mathbf{1}^T s \quad (11)$$

$$s.t. \quad Hx + Jz + K[l + (u-l) \circ \tilde{v}] - s \leq h \quad (12)$$

**Remarks**: (1) Decision variables in the **MP** include $l, u, x^{(k)}$ and $z^{(k)}$ ($\forall k \in K$), while $\tilde{v}^{(k)}$ is an scenario identified by solving **SP**. $x^{(k)}, z^{(k)}$, and $\tilde{v}^{(k)}$ ($\forall k \in K$) are generated progressively together with constraints in (10) as iterations proceed.

(2) The **MP** is a mixed-integer linear program that can be readily solved by off-the-shelf solvers.

(3) By taking dual of the inner most minimization problem in (11), the **SP** can be transformed into a max-min-max problem, i.e. **SP-d** in the following. The **SP-d** can be effectively solved using the column-and-constraint generation (C&CG) approach [4].

$$\textbf{(SP-d)} \quad Q(l,u) = \max_{\tilde{v} \in [0,1]^n} \min_{z \in \mathbb{N}} \max_{0 \leq \lambda \leq 1} \lambda^T[K(l + (u-l) \circ \tilde{v}) + Jz - h] \quad (13)$$

$$s.t. \quad \lambda^T H = 0 \quad (14)$$

## IV. CASE STUDIES

We conduct case studies on a modified IEEE 118-bus system. The system consists of 71 conventional thermal units, 4 QSUs, 5 wind farms, and 186 transmission lines. Detailed data of the test system are provided in [5].

The DNE limits of wind farms for $T = 4$ ($\Delta T = 5$ min) periods are calculated in a two-step framework [1]: In the first step, dynamic economic dispatch is performed to provide desired dispatch points (DDPs) of conventional units. The resulting locational marginal prices (LMPs) are used to set the coefficients $\sigma$ [1]. In the second step, based on the DDPs, the DNE limits of wind farms are calculated using the proposed model. We consider three cases: Case 1: No commitments of QSUs are adjustable for recourse controls; Case 2: Commitments of two QSUs are adjustable for recourse controls; Case 3: Commitments of four QSUs are adjustable for recourse controls.

Fig. 1 shows the DNE limits of total wind output in all cases. We observe that the total wind power dispatchable ranges in Cases 2 and 3 with DRCs are wider than those in Case 1 without DRCs. The adjustment of QSU commitments increases the system's regulation capability to respond to wind power volatility. Therefore the maximum allowable ranges of VRG output are broadened. In addition, more available DRCs could further improve the operation flexibility to accommodate wind power generation, as observed in Fig. 1.

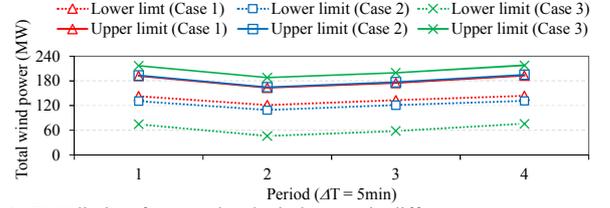

Fig. 1. DNE limits of system-level wind output in different cases.

For Case 3, another set of DNE limits is obtained by solving a single-period DNE limit model at each of the four periods. As shown in Fig. 2, the single-period dispatchable ranges are wider than the multi-period ones, because inter-temporal constraints are not considered in the single-period model. However, the single-period ranges are over-optimistic, within which corrective dispatch may not be feasible in successive periods. The red curve in Fig. 2 represents a scenario that is within the single-period dispatchable ranges but out of the multi-period ones. It is verified that this scenario leads to infeasible corrective dispatch as ramping resources are insufficient. This result shows the necessity of involving multiple periods in DNE limits.

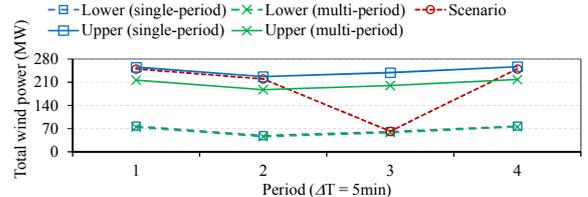

Fig. 2. DNE limits of system-level wind output by different approaches.

## V. CONCLUSIONS

A multi-period DNE limit model with continuous and discrete recourse controls is proposed. The proposed model is formulated as a two-stage RO with mixed-integer recourse. The NC&CG method is used to solve the proposed model exactly. Simulation results show that the proposed method manages to improve the DNE limits by employing the additional flexibility of DRCs and ensure the feasibility of corrective dispatch in successive periods.